# Sparse Optimization of Two-Dimensional Terahertz Spectroscopy

Z. Wang[1,2,3], H. Da[4], A. S. Disa[5], T. Pullerits[6], A. Liu[7,*], F. Schlawin[1,2,8,*]

[1]Max Planck Institute for the Structure and Dynamics of Matter, Hamburg, Germany
[2]University of Hamburg, Luruper Chaussee 149, Hamburg, Germany
[3]Department of Physics, Stockholm University, Albanova University Centre, SE-106 91 Stockholm, Sweden
[4]CNPC Research Institute of Safety and Environment Technology, Beijing 102206, China
[5]School of Engineering and Applied Physics, Cornell University, New York, USA
[6]Department of Chemical Physics, Lund University, P.O. Box 124, Lund 22100, Sweden
[7]Condensed Matter Physics and Materials Science Division, Brookhaven National Laboratory, New York, USA
[8]The Hamburg Centre for Ultrafast Imaging, Hamburg, Germany

*Corresponding authors: aliu1@bnl.gov; frank.schlawin@uni-hamburg.de

**Two-dimensional terahertz spectroscopy (2DTS) is a low-frequency analogue of two-dimensional optical spectroscopy that is rapidly maturing as a probe of a wide variety of condensed matter systems. However, a persistent problem of 2DTS is the long experimental acquisition times, preventing its broader adoption. A potential solution, requiring no increase in experimental complexity, is signal reconstruction via compressive sensing. In this work, we apply the sparse exponential mode analysis (SEMA) technique to 2DTS of a cuprate superconductor. We benchmark the performance of the algorithm in reconstructing the terahertz nonlinearities and find that SEMA reproduces the asymmetric photon echo lineshapes with as low as a 10% sampling rate and reaches the reconstruction noise floor with beyond 20-30% sampling rate. The success of SEMA in reproducing such subtle, asymmetric lineshapes confirms compressive sensing as a general method to accelerate 2DTS and multidimensional spectroscopies more broadly.**

Multidimensional coherent spectroscopies [1, 2, 3] have revolutionized our understanding of complex systems ranging from molecular liquids [4, 5, 6, 7] to quantum-confined nanostructures [8, 9, 10], and even biological complexes [11, 12, 13, 14]. In recent years, two-dimensional terahertz spectroscopy (2DTS) [15, 16] has brought the unique capabilities of multidimensional techniques to condensed matter systems [17], in which many fundamental excitations can be found at low energies [18, 19]. Recent such experiments have studied, for example, ferroelectrics

[20, 21], ferromagnets [22, 23], and even superconductors [24, 25]. However, the technique of 2DTS is still in a nascent stage, with insufficient acquisition efficiencies remaining an obstacle to studying materials with small nonlinear optical signals.

At optical and infrared frequencies and in nuclear magnetic resonance, there has been tremendous effort in accelerating multidimensional spectroscopic techniques [26, 27, 28]. Yet the need to accelerate 2DTS is even more pressing, since unique challenges such as long data acquisition (with reported acquisition times reaching one week for a single spectrum [29]) and potential degradation of terahertz generation over time [30] restrict the range of applications. Currently, the primary method for accelerating 2DTS is single-shot THz detection [31], where the entire THz waveform is captured simultaneously. There are various methods to implement single-shot detection schemes [32], but all of them inevitably increase experimental complexity and have their unique trade-offs. Other methods of accelerating 2DTS are therefore desirable.

In contrast to increasing signal acquisition rate, an alternative approach to accelerating 2DTS is to reduce the requirements for signal acquisition itself. For a signal sampled uniformly in time, it is well-known [33] that the Nyquist criterion requires a minimum sampling rate of twice the signal frequency. However, one may circumvent this limit by non-uniform sampling and subsequent signal reconstruction via compressive sensing algorithms [34, 35, 36, 37]. So far, compressive sensing has been successfully demonstrated not only in ultrafast spectroscopy [38], but also in multidimensional NMR [39] and multidimensional optical spectroscopies [40, 41, 42, 43]. However, these techniques have yet to be applied towards 2DTS, which stands to benefit even more from acceleration. Compressive sensing has also not been applied to asymmetric two-dimensional spectral lineshapes, which are frequently encountered in disordered systems [44, 45] and strong vibronic coupling [46, 47] more generally. Here, we address these two open problems and implement the sparse exponential mode analysis (SEMA) method based on dictionary learning, which has been described in detailed elsewhere [41, 42]. We choose the SEMA method as it not only requires fewer data points than traditional compressive sensing methods such as LASSO and Matching Pursuit, but also because it can reconstruct both resonance frequencies and linewidths simultaneously.

As an ideal test case, we apply SEMA towards reconstructing 2DTS spectra of the Josephson plasma resonance [48] in the optimally-doped cuprate superconductor La$_{1.83}$Sr$_{0.17}$CuO$_4$ (LSCO), which has a plasma frequency $f_p = 2$ THz. The experiment is shown schematically in Figure 1a, in which two terahertz excitation pulses ($E_A$ and $E_B$) polarized along the c-axis of LSCO drive interlayer supercurrents that radiate a nonlinear optical signal $E_{NL}$ as a function of inter-pulse time delay $\tau$ and laboratory time t. In particular, we measure the 'Josephson echo' signal that has previously been used [25] to measure disordered superconductivity in this same compound. The underlying wavevector phase-matching condition that isolates the echo signal has been described elsewhere [25].

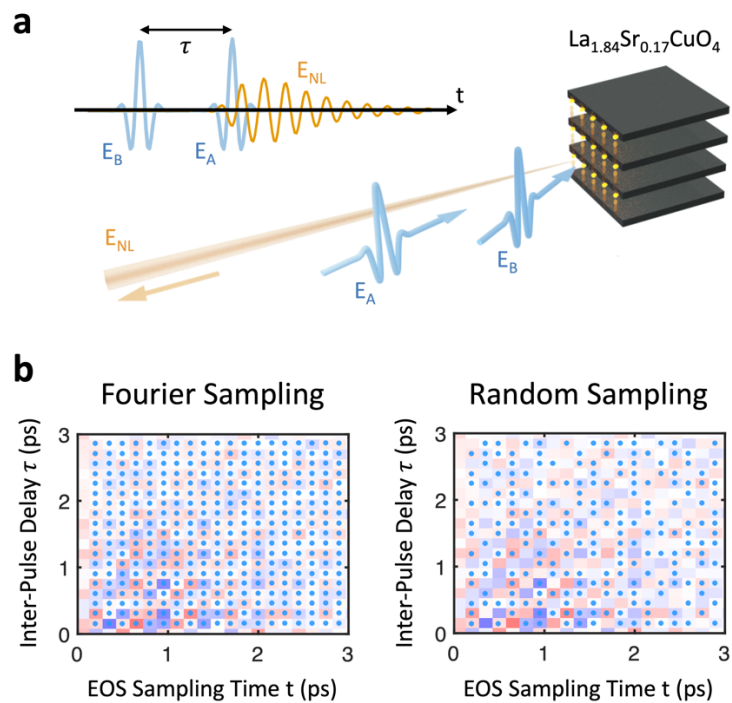

**Fig. 1. Random sampling in two-dimensional terahertz spectroscopy. a.** Schematic of the 2DTS measurement, in which two excitation fields $E_A$ and $E_B$ cooperatively drive nonlinearities of the Josephson plasma resonance in optimally-doped La$_{1.84}$Sr$_{0.17}$CuO$_4$. The resulting supercurrents radiate a nonlinear electric field $E_{NL}$, which is measured as a function of the inter-pulse time delay $\tau$ and the laboratory time t as shown inset. **b.** Two possible acquisition schemes of the nonlinear signal $E_{NL}$, where sampled data points are indicated by the blue dots. (Left) A uniform sampling grid appropriate for Fourier transform into the frequency-domain. (Right) Non-uniform sampling appropriate for reconstruction via compressive sensing algorithms.

In Figure 1b, we describe the two time-domain acquisition schemes for generating a 2DTS spectrum. To the left we depict conventional Fourier sampling of $E_{NL}$ with a uniform sampling grid, where the time-step determines the frequency bandwidth and the time-range determines the frequency resolution, respectively. To the right we depict sparse sampling of $E_{NL}$ where the signal is randomly sampled across the same temporal range, generally with far fewer data points. In this case, as shown by Tao, Romberg, and Candes [35], the Nyquist criterion may be circumvented using appropriate reconstruction algorithms.

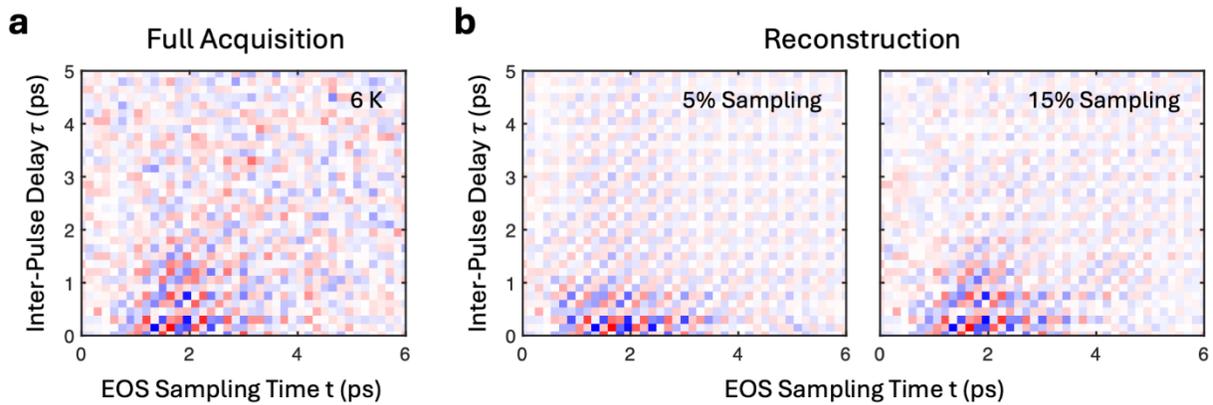

**Fig. 2. Time-domain compressive sensing. a.** Fourier-sampled Josephson echo signal (at a sample temperature of 6 K) in the time-domain. **b.** Reconstructions of the Josephson echo signal at the same time coordinates $\{\tau, t\}$ from sparse-sampling of 5% and 15% of the data points in **a**. More accurate reconstruction is evident with increasing sampling percentage.

We begin by measuring the Josephson echo signal at a temperature of 6 K via Fourier sampling, as shown in Figure 2a. The signal is sampled with a time range of 6.9 ps and 7.95 ps along $\tau$ and $t$ respectively with identical time steps of 150 fs, resulting in 46 x 53 = 2438 total sampled data points before zero-padding by twice the sampling size [7] and a total acquisition time of approximately 3 hours. We then use the SEMA algorithm to reconstruct the signal, which fits a sparsely-sampled dataset to a dictionary of frequencies and spectral linewidths. This dictionary is refined iteratively, until convergence is reached [41, 42]. To investigate the accuracy of the SEMA reconstruction method, we non-uniformly sample a fraction of the Fourier-sampled dataset as input to our compressive sensing algorithm for subsequent reconstruction. The reconstructed time-domain signal is shown in Figure 2b for sampling percentages of 5% and 15% of the original dataset, which exhibit qualitative differences. While the reconstruction with 5% of the data can be seen to reproduce the oscillation frequencies and their relative phase along each time axis, the reconstruction of 15% more accurately reproduces the oscillation lifetimes of the Josephson echo signal. To more easily infer the reconstruction accuracy, it is instructive to examine the spectral lineshapes of the Josephson echo signal in the frequency-domain.

In Figure 3, we show the 2DTS spectra obtained by Fourier transform of the time-domain data in Figure 2 into the frequency-domain. Fourier transform of the original Fourier-sampled signal (with zero-padding by twice the sampling size) returns the asymmetric "almond-shaped" peak shown in Figure 3a, indicative of a disordered Josephson plasma resonance as discussed in [25]. In comparison, Fourier transform of the reconstructed time-domain data reveals a strong dependence on the percentage of data used for reconstruction. The reconstruction of 5% exhibits a cross-shaped peak typical of a homogeneously-broadened resonance [44] while the reconstruction of 15% qualitatively reproduces the true asymmetric lineshape.

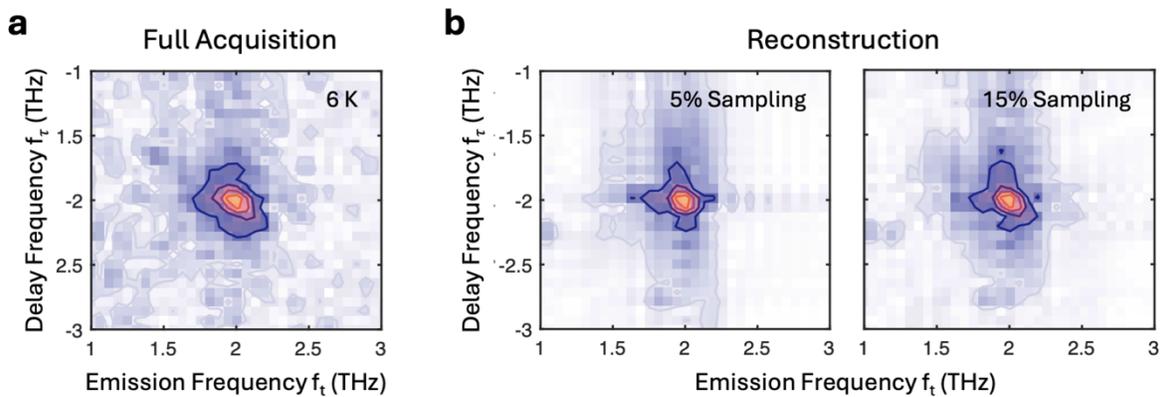

**Fig. 3. Frequency-domain compressive sensing. a.** Reference 2DTS spectrum acquired by Fourier transform of the Fourier sampled time-domain data in Figure 2a. **b**. Reconstructed 2DTS spectra acquired by Fourier transform of the reconstructed time-domain data in Figure 2b. At 5% sampling percentage a symmetric peak is observed, while increasing sampling percentage retrieves the asymmetric 'echo' lineshape.

We now examine the accuracy of the reconstructed two-dimensional lineshapes more closely. Slices of both the Fourier-sampled and reconstructed 2DTS spectra are taken along the 'diagonal' ($|f_t| = |f_\tau|$) and perpendicular 'cross-diagonal' directions used to characterize intrinsic and disorder broadening [44]. Comparison of slices taken of the full Fourier-sampled spectrum and those of the 5% reconstruction are shown in Figure 4a, where we see that the reconstruction is less accurate and returns similar linewidths in both directions. The reconstruction further misses the non-Lorentzian tails of the resonance entirely. However, as shown in Figure 4b, increasing the sampling percentage to 15% results in a reconstruction that accurately reproduces not only both linewidths, but even subtle details of the fully-sampled spectral lineshapes.

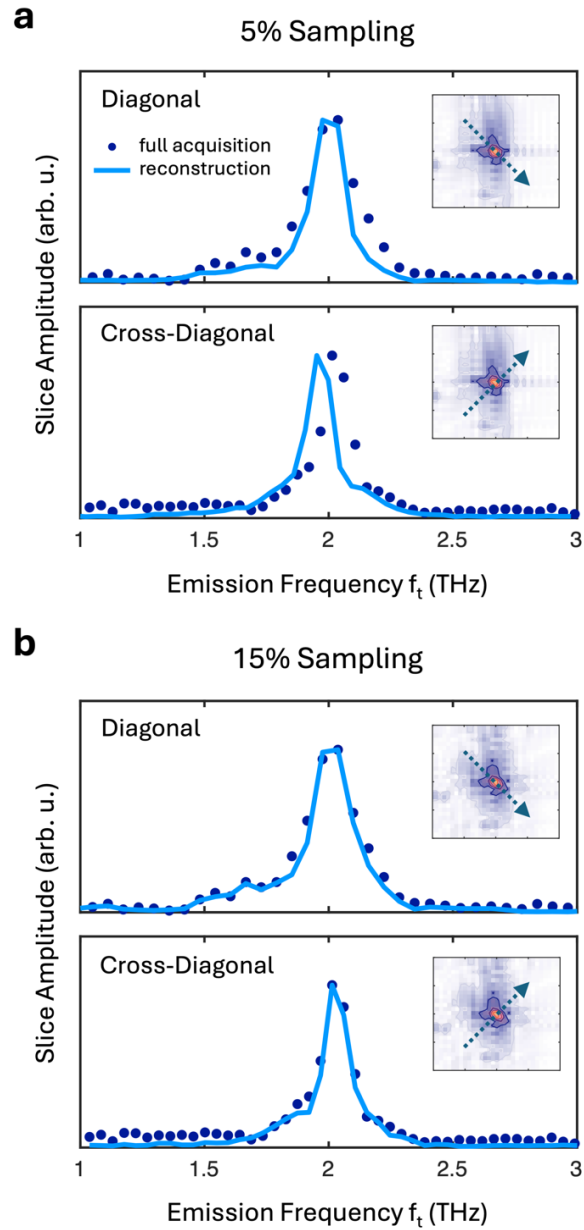

**Fig. 4. Reconstructed lineshapes.** Comparison of cross-sections taken (indicated inset) from the original reference 2DTS spectrum to those taken from the 2DTS spectra reconstructed from **a**. 5% sampling percentage and **b**. 15% sampling percentage.

Finally, we quantify the reconstruction accuracy by considering the residual error of each reconstruction. We obtain residual maps of the 2DTS spectra by subtracting the reconstructed spectra from the fully-sampled spectra, which are then normalized to the maximum amplitude of the fully-sampled spectra and plotted in Figure 5a. We note that the residual map of the 5% sampling spectrum exhibits significant structure, with positive (red) error near the peak center and negative (blue) error in the wings of the peak. However, the residual map of the 15% sampling spectrum exhibits much weaker structure with only the negative (blue) error in the wings remaining apart from the reconstruction noise. This is supported by the integrated residual error shown in Figure 5b, which approaches the noise floor with increasing sampling percentage as expected. Beyond 30% sampling percentage, the reconstruction accuracy at this sampling percentage is primarily limited by the measurement signal-to-noise ratio.

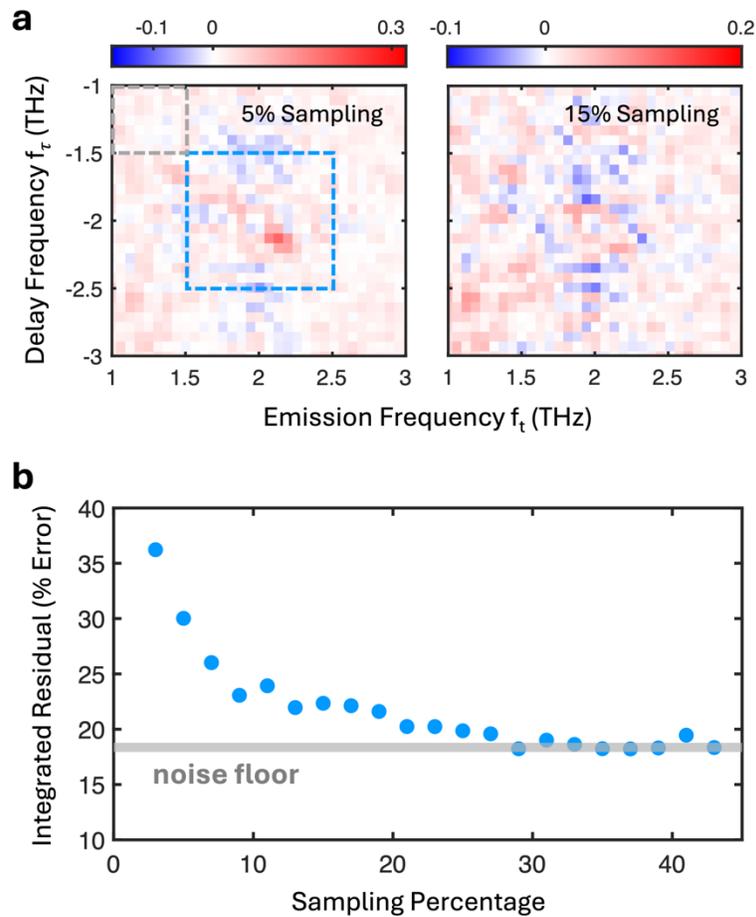

**Fig. 5. Reconstruction residual error. a.** Residual error of the reconstructions found by subtracting the reference spectrum from the reconstructed spectra at sampling percentages of 5% and 15% and normalizing to the maximum amplitude of the reference spectrum. **b**. Integrated error found by summing the (unnormalized) magnitude of the residual error across the region marked by the dashed blue box in the left panel of **a** and normalizing to the integrated value of the reference spectrum across the same frequency range. The indicated reconstruction noise floor is found by performing the same procedure for the region marked by the dashed gray box in the left panel of **a**, in which no signal is present, and scaling by an area factor of 4.

Having demonstrated compressive sensing as a viable method to reconstruct 2DTS spectra, we now turn to an outlook for its application to real experimental scenarios. In the results presented here, we find an approximate lower limit of 10% sampling percentage for accurate reconstruction (5% error above the reconstruction noise floor) of the 2DTS spectrum and further reach the reconstruction noise floor above 20-30% sampling percentage. However, we emphasize that the Josephson echo signal considered here exhibits a more complicated spectral lineshape than most other 2DTS signals reported to date. For reconstructing symmetric 'non-rephasing' signals or signals from homogeneously-broadened systems more generally, we expect these limits on the sampling percentage to be relaxed even further. Finally, combining such sparse optimization with other techniques such as single-shot THz detection [31] will bring a new level of versatility to 2DTS techniques, enabling study of systems with weak optical nonlinearities and fragile systems sensitive to external perturbation. Further algorithmic developments may also yield additional performance improvements.

## Methods

**2-D Terahertz Spectroscopy**

To perform 2-D terahertz spectroscopy, two intense terahertz pulses were generated by optical rectification of 100 fs, 1300 nm pulses in two OH1 (2-{3-(4-hydroxystyryl)-5,5-dimethylcyclohex-2-enylidene}malononitrile) organic crystals. The two terahertz pulses are then focused in a non-collinear geometry onto the sample with a parabolic mirror of focal length 76.2 mm [49], resulting in peak electric fields of ~25 kV/cm and ~10 kV/cm at the sample surface. The emitted nonlinear electric field was then detected by conventional electro-optic sampling using 100 fs, 800 nm pulses in a ZnTe crystal. A differential chopping scheme, in which E$_A$ and E$_B$ were modulated at 500 Hz and 333 Hz respectively, was used to isolate the nonlinear electric field from the excitation fields.

**Compressive Sensing**

Compressive sensing theory aims to reconstruct a signal from a subset of measurements $y$ with $m$ elements. Naturally, the number of elements should be much smaller than the number of data points in the full signal, $m \ll M$. These measurements are chosen randomly from the original data. The reconstruction uses a dictionary matrix $\widetilde{A}$ of size $m \times n$ (where $m < n$), which is also referred to as the sparse matrix or sampling matrix in compressive sensing theory. The dictionary elements are chosen as $\widetilde{A}_{k,j} = e^{i(\omega_{1j} + i\beta_{1j}) \cdot t_{1k}} \otimes e^{i(\omega_{2j} + i\beta_{2j}) \cdot t_{2k}}$, with the times $t_{1k}$ and $t_{2k}$ ($k = 1..m$) extracted from the randomly selected measurements. The frequencies $\omega_{1j}$ and $\omega_{2j}$ and damping rates $\beta_{1j}$ and $\beta_{2j}$ are chosen initially to form a very rough grid around the plasma frequency (see SM for details). The reconstructed signal is then given by $\widetilde{A}\widetilde{g}$, where the weight vector $\widetilde{g}$ is a compressed, low-dimensional representation of the sparse basis. The SEMA algorithm iteratively refines the dictionary matrix. In each iteration, a convex minimization of the cost function

$$minimize_{\widetilde{g}} \left\{ \frac{1}{2} \|y - \widetilde{A}\widetilde{g}\|_2^2 + \lambda \|\widetilde{g}\|_1 \right\} \quad (1)$$

Is carried out. Here, $\lambda$ is a Lagrange coefficient which adds a penalty for solutions with large 1-norm, and thus favors solutions with the smallest number of nonzero parameters. Its size is adjusted within the iteration loop [42, 41]. For further details on the reconstruction process, please refer to the Supporting Information of [41].

## Acknowledgments

Z. W. acknowledges support from the Swedish Research Council VR (Grants 2022–06176), and to Dieter Jaksch and Markus Kowalewski for fruitful discussions. We greatly appreciate Andreas Jakobsson's previous help with theory, and Andrea Cavalleri for the use of 2DTS data measured at the Max Planck Institute of Structure and Dynamics of Matter. F. S. acknowledges support from the Cluster of Excellence 'Advanced Imaging of Matter' of the Deutsche Forschungsgemeinschaft (DFG) - EXC 2056 - project ID 390715994. The work at BNL was supported by the U.S. Department of Energy (DOE), Office of Basic Energy Sciences (BES) under contract no. DOE-SC0012704. The work at Lund University was supported by the Swedish Research council (2021-05207) and Swedish Energy Agency (50709-1).